\documentclass[letterpaper, 10 pt, conference]{ieeeconf}  

\IEEEoverridecommandlockouts                              

\usepackage{graphicx} 
\usepackage{amsmath} 
\usepackage{amssymb}  
\usepackage{footnote}
\usepackage{lipsum} 
\usepackage{multirow}
\usepackage{multicol}
\usepackage{bm}

\DeclareMathOperator*{\argmin}{arg\,min}
\DeclareMathOperator{\Tr}{Tr}

\title{\LARGE \bf
Ego-noise reduction of a mobile robot using noise spatial covariance matrix learning and minimum variance distortionless response
}

\author{Pierre-Olivier Lagacé, Fran\c{c}ois Ferland, and Fran\c{c}ois Grondin
\thanks{This work was supported by NSERC-CREATE/CoRoM and a NSERC Discovery Grant. P.-O. Lagacé, F. Ferland and F. Grondin are with the Department of Electrical Engineering and Computer Engineering, Universit\'e de Sherbrooke.}%
}

\begin{document}

\maketitle
\thispagestyle{empty}
\pagestyle{empty}

\begin{abstract}
The performance of speech and events recognition systems significantly improved recently thanks to deep learning methods.
However, some of these tasks remain challenging when algorithms are deployed on robots due to the unseen mechanical noise and electrical interference generated by their actuators while training the neural networks.
Ego-noise reduction as a preprocessing step therefore can help solve this issue when using pre-trained speech and event recognition algorithms on robots.
In this paper, we propose a new method to reduce ego-noise using only a microphone array and less than two minute of noise recordings.
Using Principal Component Analysis (PCA), the best covariance matrix candidate is selected from a dictionary created online during calibration and used with the Minimum Variance Distortionless Response (MVDR) beamformer.
Results show that the proposed method runs in real-time, improves the signal-to-distortion ratio (SDR) by up to 10 dB, decreases the word error rate (WER) by 55\% in some cases and increases the Average Precision (AP) of event detection by up to 0.2.
\end{abstract}

\section{INTRODUCTION}
In human-machine interaction, the audio modality plays a central role in speech and sound-event recognition.
In the last few years, robust speech and event recognition based on transformers have been proposed (e.g. Whisper \cite{radford2022robust}, Google Speech-to-Text \cite{chiu2018state} and PANNS \cite{kong2020panns}).
Although these systems perform well with ambient noise, their performance drops quickly with unusual noises and low Signal-to-Noise Ratio (SNR). 
When microphones are mounted on a robot, the audio signal quality decreases considerably due to noise produced by the motors and WiFi/Bluetooth antennas. 
This makes denoising challenging, as the noise generated by the robot, referred to as ego-noise, is usually non-stationary.

Numerous approaches have been proposed to reduce the ego-noise for Unmanned Ground Vehicles (UGVs) and Unmanned Aerial Vehicles (UAVs). 
Time-Frequency Spatial filtering (TFS) \cite{wang_tfs_2017, wang_acoustic_2018} uses the direction of arrival (DOA) of the target sound source at each time-frequency bin to mask interfering noise.
Since it relies on spatial cues, performance decreases when the ego-noise and target sound are close.
On the other hand, Blind Source Separation (BSS) \cite{wang_ear_2016, wang_microphone-array_2017} extracts each sound source individually using Independent Component Analysis (ICA) and permutation alignment.
BSS works efficiently with low Signal-to-Noise Ratios (SNRs), but suffers from the permutation ambiguity.
A TFS-BSS approach \cite{wang_blind_2020} is proposed to solve the permutation ambiguity, but it relies on long audio segments, which makes it unsuitable for real-time applications.

Another strategy consists in using a reference signal captured by a microphone \cite{kang_software_2019, yoon_advanced_2015, yoon_two-stage_2016} or a piezoelectric sensor \cite{fernandes_first_2015} close to the motors.
This assumes the target sound is not captured by the reference microphone.
However, this depends on the robot configuration, which makes it difficult to generalize to any robots.
Other methods rely on a database that contains noise templates and motor inertial data to select the best noise instance that fits the observed actuator profile \cite{ince_online_2012, ince_ego_2009, ince_hybrid_2010, ince_incremental_2011}.
However, this requires synchronization between audio and inertial signals. 
A similar method consists in using a audio only dictionary-based approach. 
The dictionary is trained to represent the multiple components of the ego-noise, and then used to reconstruct and suppress the ego-noise from the noisy signal. 
Non-Negative Matrix Factorization (NMF)  \cite{schmidt_informed_2019, schmidt_motor_2020, schmidt_multichannel_2021} and phase optimized singular value decomposition (PO-KSVD) \cite{deleforge_phase-optimized_2015} can be used to train the dictionary.
The motors state are use to reconstruct the ego-noise from the dictionary components.
Some NMF approaches do not use motors state in their framework \cite{tezuka_ego-motion_2014, haubner_multichannel_2018}, which can introduce non-linear distortion and hurt speech recognition performances.
On the other hand, Fang et al. \cite{fang_joint_2021} use linear operation that does not introduce distortion, but performs poorly on non-stationary noise.
Deep-learning based methods can also be used to separate target speech from ego-noise \cite{ito_internal_2005, briegleb_deep_2019, kim_mimo_2021, tan_efficient_2019, wang_deep_2020}. 
 These models perform well when trained on large datasets that contain ego-noise and target sounds.
 However, these trained networks usually perform poorly in a different unseen audio scene, which is common in robotics.

In this paper, a template-based approach using ego-noise spatial covariance matrix (SCM) with minimum variance distortionless response (MVDR) algorithm is proposed.
The approach improve SDR while being robust to non-stationnary noise and can be quickly calibrate in novel audio domains.
This is appealing as it avoids retraining deep neural network, which usually requires hours of audio samples and computing resources not available on a robot.
A calibration step is used to generate a dictionary of ego-noise SCMs. 
Principal Component Analysis (PCA) is used to reduce the number of dimensions of the SCM. 
This compact representation is used to find the best fit between the SCMs in the database and the measured SCM. 
The corresponding SCM is used in the MVDR algorithm to separate the target sound source from the ego-noise. 
Results show that this method increases both the performances of speech and event recognition.

This paper is organized as follows. Section \ref{sec:mvdr_scm} presents the overall ego-noise reduction method. Section \ref{sec:experimental_setup} describes the experimental setup, followed by Section \ref{sec:results} with the results obtained with real-recorded data and Section V with the conclusion of this paper.

\section{MVDR using SCM estimation}
\label{sec:mvdr_scm}

The multi-channel target sound, ego-noise and audio mixture for microphones $m \in \{1, \dots, M\}$, where $M \in \mathbb{N}$ stands for the number of microphones, are denoted by $\mathbf{s}[n] \in \mathbb{R}^M$, $\mathbf{b}[n] \in \mathbb{R}^M$ and $\mathbf{x}[n] \in \mathbb{R}^M$, respectively, where $n \in \mathbb{N}$ corresponds to the sample index.
It is assumed that noise is purely additive, such that:

\begin{equation}
\mathbf{x}[n] = \mathbf{s}[n] + \mathbf{b}[n].
\end{equation}

These signals can be represented in the frequency domain using a Short-Time Fourier Transform (STFT) with frame size of $N \in \mathbb{N}$ samples and hop size of $\Delta N \in \mathbb{N}$ samples, and are denoted as $\mathbf{X}[k,l] \in \mathbb{C}$ , $\mathbf{S}[k,l] \in \mathbb{C}$ and $\mathbf{B}[k,l] \in \mathbb{C}$, where $k \in \{ 0, 1, \dots, K-1 \}$ and $l \in \mathbb{N}$ represent the frequency bin and frame indices, respectively.
The spatial covariance matrices of the input signal $\Phi_{\mathbf{XX}}[k]$ is computed over $L$ frames as follows:

\begin{equation}
\Phi_{\mathbf{XX}}[k] = \frac{1}{L}\sum\limits_{l=1}^{L}{\mathbf{X}[k,l]\mathbf{X}[k,l]^H},
\label{eqn_scm}
\end{equation}
where $\{\dots\}^H$ stands for the Hermitian operator.

The calibration stage begins by recording the robot's ego-noise during 
a few minutes (in this paper, the method requires only 90 sec of ego-noise data).
The ego-noise recording is separated in $J \in \mathbb{N}$ segments of 0.5 sec, and the SCMs are computed for each segment $j$ using (\ref{eqn_scm}). 
Since the SCM is hermitian, its upper triangle (including the diagonal) can be flattened to create a vector $\mathbf{v}_j[k] \in \mathbb{C}^{M(M+1)/2}$ for each frequency bin $k$ that holds all relevant elements.
These vectors are concatenated in a supervector $\mathbf{V}_j \in \mathbb{C}^{KM(M+1)/2}$ for all frequency bins. 
Each supervector can be reduced to a dense vector using Principal Component Analysis (PCA), and denoted $\mathbf{D}_j \in \mathbb{C}^{I}$. 

At test time, the same PCA transformation is applied to the supervector obtained from the SCMs for noisy audio segments of 0.5 sec, and the result is denoted as $\hat{\mathbf{D}} \in \mathbb{C}^I$.
The $l^2$-norm is computed $\hat{\mathbf{D}}$ and vectors $\mathbf{D}_j$ from the dictionary.
The closest dictionary vector $\mathbf{D}_{j^*}$ is then selected:
\begin{equation}
\label{eqn_jmin}
    j^* = \argmin_j(\lVert \hat{\mathbf{D}} - \mathbf{D}_j\rVert^2_2).
\end{equation}

Dimensionality reduction with a PCA serves two purposes: 1) reduction of the memory footprint and number of computations; 2) projection of the supervector observation on the noise subspace, which eases comparison between the noisy SCMs (target and noise) and the noise SCMs.
PCA decomposition can be computed efficiently on a 
low-power embedded computer, which makes dictionary generation on a robot fast once the calibration audio signals are recorded. 
The corresponding noise SCM estimate $\hat{\Phi}_{\mathbf{BB}}[k]$ is used to obtain the MVDR weights:
\begin{equation}
\label{eqn_w}
\mathbf{w}[k] = \frac{\hat{\Phi}^{-1}_{\mathbf{BB}}[k]\Phi_{\mathbf{XX}}[k]}{\Tr\{\hat{\Phi}^{-1}_{\mathbf{BB}}[k]\Phi_{\mathbf{XX}}[k]\}} \mathbf{u},
\end{equation}
where the inverse of each SCM noisy matrix (denoted as $\hat{\Phi}^{-1}_{\mathbf{BB}}[k]$) can be precomputed once during calibration and directly stored in the dictionary to speed up computations.
The one-hot reference vector $\mathbf{u} \in \{0, 1\}^M$ is chosen to select the microphone with the highest SNR, which can be estimated from the selected noise covariance matrix.
Beamforming is finally applied and generates the enhanced signal:

\begin{equation}
\label{eqn_Y}
Y[k,l] = \mathbf{w}[k]^{H}\mathbf{X}[k, l].
\end{equation}

The inverse STFT (iSTFT) can then be used to transform back the signal to the time domain and obtain $y[n]$.

\begin{figure}[!ht]
\vspace{5pt}
\centering
\includegraphics[width=\linewidth]{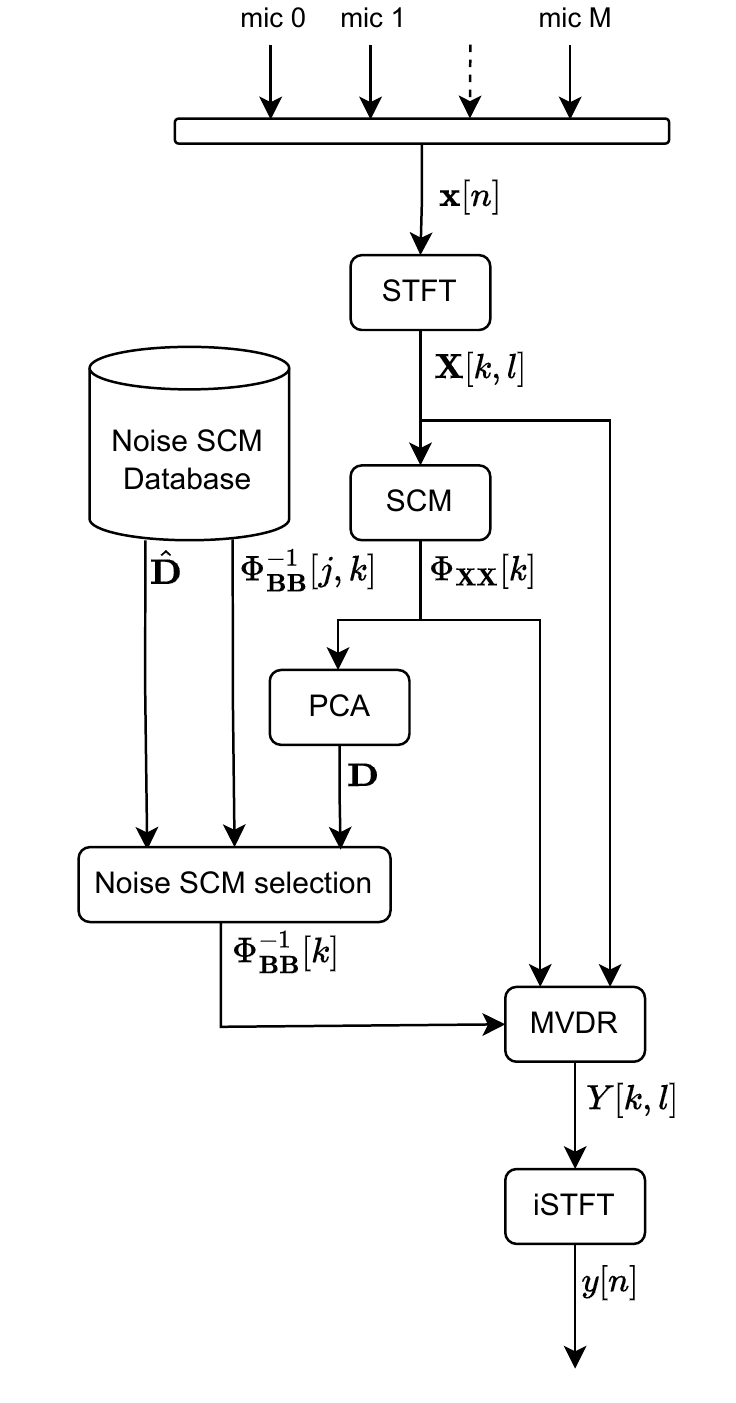}
\caption{Block diagram of the proposed framework.}
\vspace{-10pt}
\label{fig_system}
\end{figure}

Figure \ref{fig_system} summarizes the proposed framework. 

\section{Experimental setup}
\label{sec:experimental_setup}

\begin{figure}[!h]
\centering
\includegraphics[width=\linewidth]{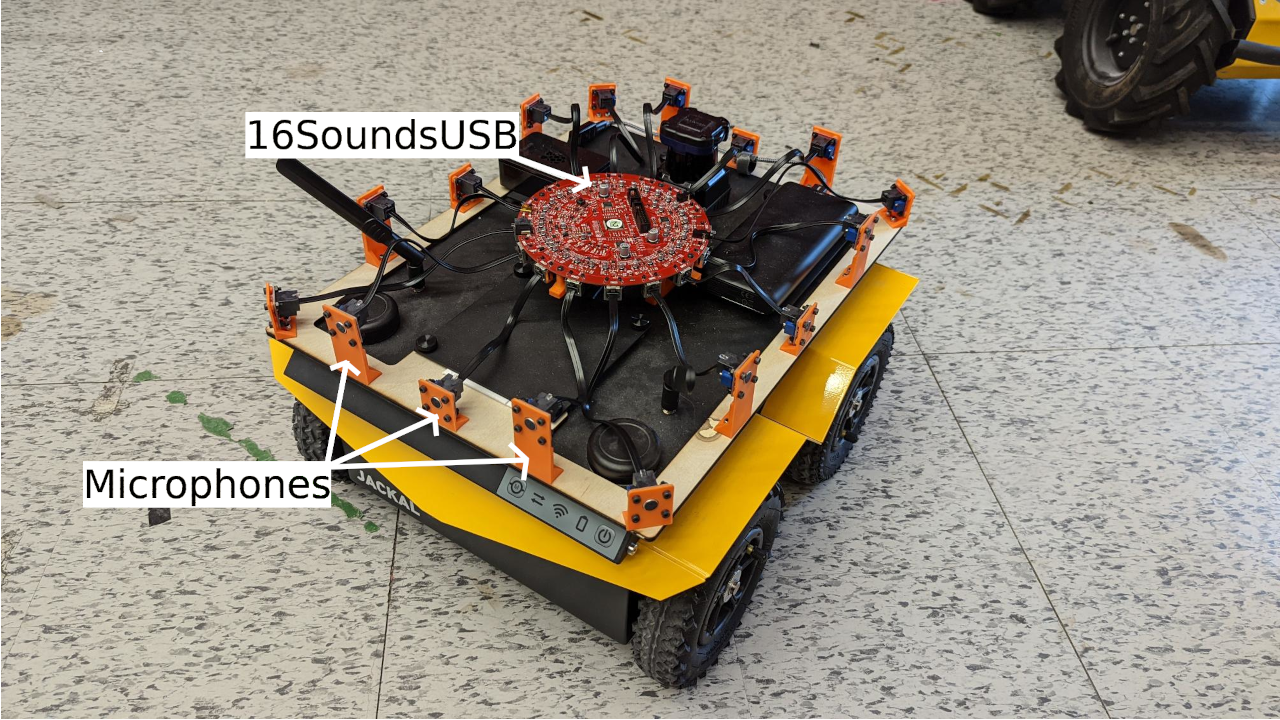}
\caption{Clearpaths Jackal Robot equiped with the microphone array}
\label{fig_robot_setup}
\end{figure}

Figure \ref{fig_robot_setup} shows the small Clearpath Robotics Jackal UGV\footnote{https://clearpathrobotics.com/jackal-small-unmanned-ground-vehicle/} equiped with a 16SoundsUSB\footnote{https://github.com/introlab/16SoundsUSB} MA used in this experiment. 
The 16 omnidirectionnal microphones are positioned around the robot frame. Half of them are 3 cm higher than the others, to ensure spatial discrimination in the $z$-axis. 
The MA is connected by USB to a laptop computer which runs the framework implemented on Python and ROS\footnote{https://github.com/introlab/egonoise/}. 
Audio is sampled at 32000 samples/sec and divided in segments of 0.5 sec.
The STFT uses a Hann window of $N=2048$ samples and a hop size of $\Delta N = 256$ samples.

The experiments are conducted in three different rooms: a small office room of $30$ m\textsuperscript{2}, a large room of $150$ m\textsuperscript{2} and a 
hallway of $30$ m\textsuperscript{2}. 
The reverberation time $RT_{60}$ are respectively 200 msec, 600 msec and 160 msec, and estimated from recorded hand claps in each room. 

For evaluation purposes, ego-noise and speech are recorded separately. 
Ego-noise is recorded in each room while the robot moves in the room at different speeds (up to 0.4 m/sec), and in arbitrary directions. 
Figure \ref{fig_robot_setup} shows the robot moving in the large room.  
While recording ego-noise, no other sound sources are active around the robot except ambient noises. 
A loudspeaker then plays target sound sources, with the volume adjusted to a casual conversation level. 
The loudspeaker is moved manually around the robot at a distance ranging from 1 and 3 meters to recreate relative motion between the robot and the target sound source. 
The target sound sources consist of speech, music,  screaming, alarms and door slams. 
A total of 36 audio files from the LibriSpeech \cite{panayotov2015librispeech} dataset are used, from 3 male and 3 female speakers. 
These speech clips vary in duration between 6 and 15 sec. 
The loudspeaker also plays music (5 sec), screaming (5 sec) and alarms sounds (3 sec). 
Sound from the slamming door comes from the actual door in the room.
A total of 90 sec of ego-noise is recorded for calibration, and another 120 sec is recorded and  mixed randomly with target sound source recordings for validation purposes. 
This provides 288 noisy speech clips and 120 noisy sound events for each room.

\begin{figure}
\vspace{5pt}
\centering
\includegraphics[width=\linewidth]{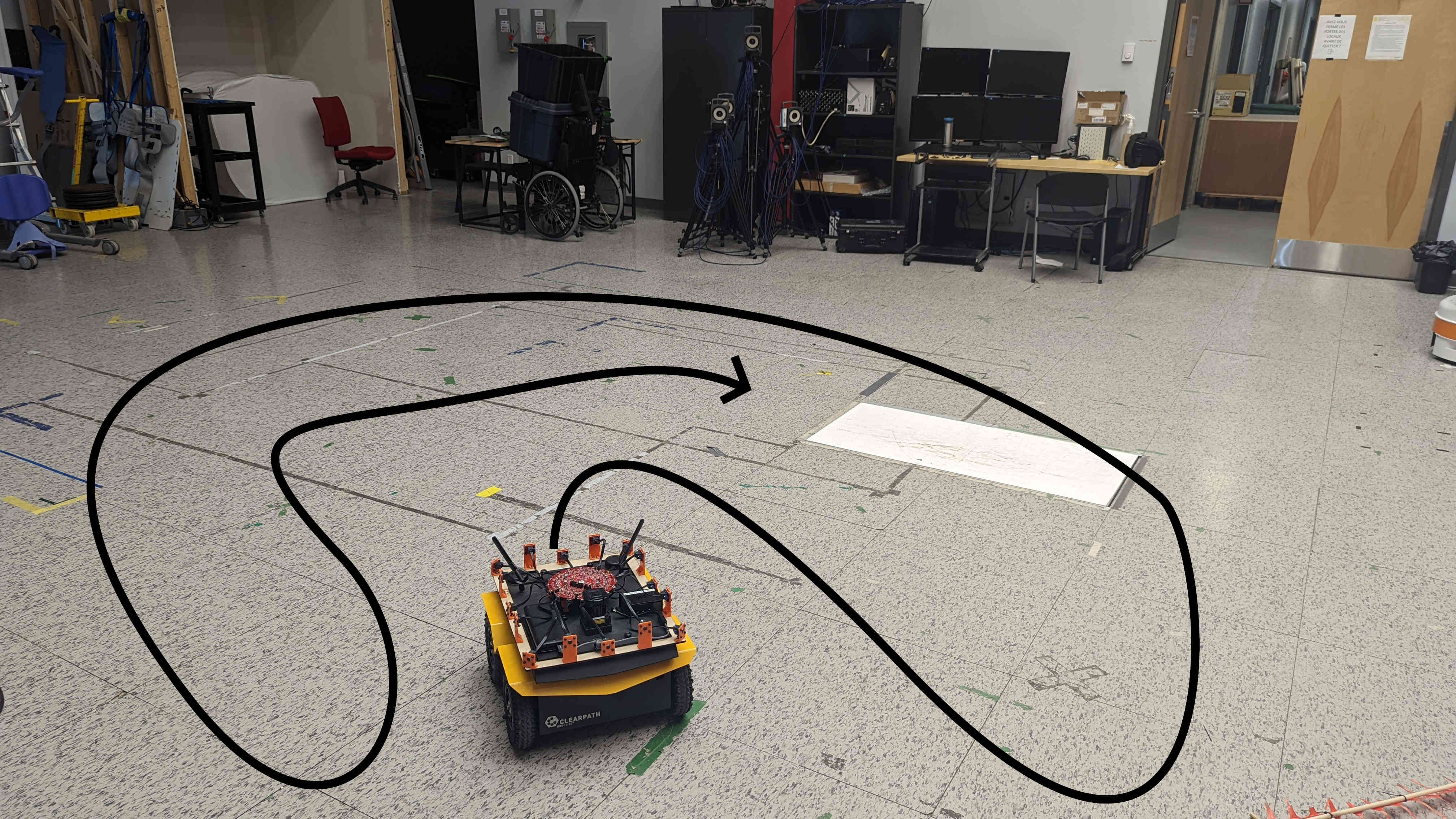}
\caption{Robot moving in the large room}
\label{fig_moving_robot}
\end{figure}

\section{Results and discussion}
\label{sec:results}

The metrics used to analyse the performance of the proposed framework is the Signal-to-Noise Ratio (SNR), the Signal-to-Distortion Ratio (SDR), the Word Rate Error\footnote{https://torchmetrics.readthedocs.io} (WER) for speech recognition and the Average Precision\footnote{https://scikit-learn.org} (AP) of classification for events detection. 
For Speech recognition, the small English model from Whisper is used. 
For sound events classification, the Wavegram Logmel Cnn14 model from PANNS is used. 
For each metric, the enhanced signal from the framework is compared to the input and voice only signals. 
The input and voice only signals are selected from one microphone in the MA which shows on average a superior SNR value compared to some other microphones closer to the actuators.

\subsection{Speech recognition} 
\label{sec:results_speech}

\begin{figure}
    \centering
    \includegraphics[width=\linewidth]{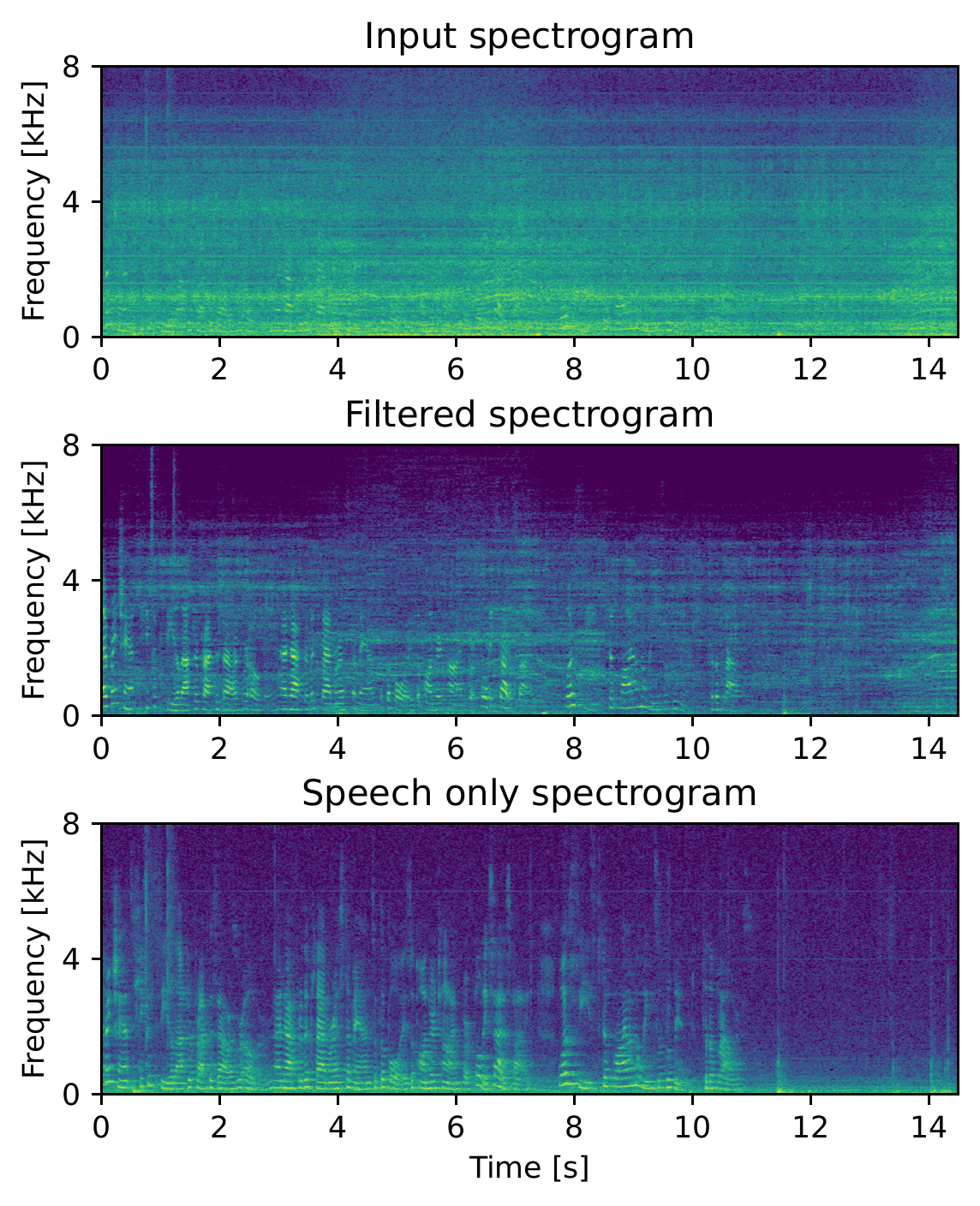}
    \vspace{-20pt}
    \caption{Spectrograms on the input, filtered and voice only signal. The SNR of the input and filered signal are -6.09 dB and 6.31 dB, respectively, which means an improvement of 12.4 dB.}
    \label{fig:specs}
\end{figure}

Figure \ref{fig:specs} shows an example of the input, filtered and voice only spectrograms in the large room with an input SNR of -6.09 dB.  
In the input spectrogram, the speech is almost entirely masked by the ego-noise. 
The filtered signal using the proposed approach has a SNR of 6.31 dB and numerous speech features now show on the spectrogram.
A Dell XPS laptop can process a 0.5 sec audio segment in 0.2 sec in the Python environment using Numpy \footnote{https://numpy.org}, which confirms that the framework can perform ego-noise reduction in real time.

Table \ref{table_snrsdr} presents the average SNRs and SDRs for each room for speech segments. 
For these metrics, only segments containing the target sounds are used and silence segments are discarted.

\begin{table}
\caption{Average input and enhanced SNRs and SDRs for each room for speech segments}
\vspace{-10pt}
\label{table_snrsdr}
\begin{center}
\renewcommand{\arraystretch}{1.5}
\begin{tabular}{c c c c c}
\hline
\hline
\multirow{2}{*}{\textbf{Room}} & \multicolumn{2}{c}{\textbf{SNR (dB)}} & \multicolumn{2}{c}{\textbf{SDR (dB)}} \\
& \textbf{Input} & \textbf{Enhanced} & \textbf{Input} & \textbf{Enhanced} \\
\hline
Small & 1.5 & 11.29 & 2.14 & 11.87 \\
Large & -2.69 & 8.7 & -1.75 & 9.57 \\
Hallway & 1.6 & 12.06 & 2.28 & 12.72 \\
\hline
\hline
\end{tabular}
\end{center}
\vspace{-15pt}
\end{table}

\begin{figure}
\centering
\includegraphics[width=\linewidth]{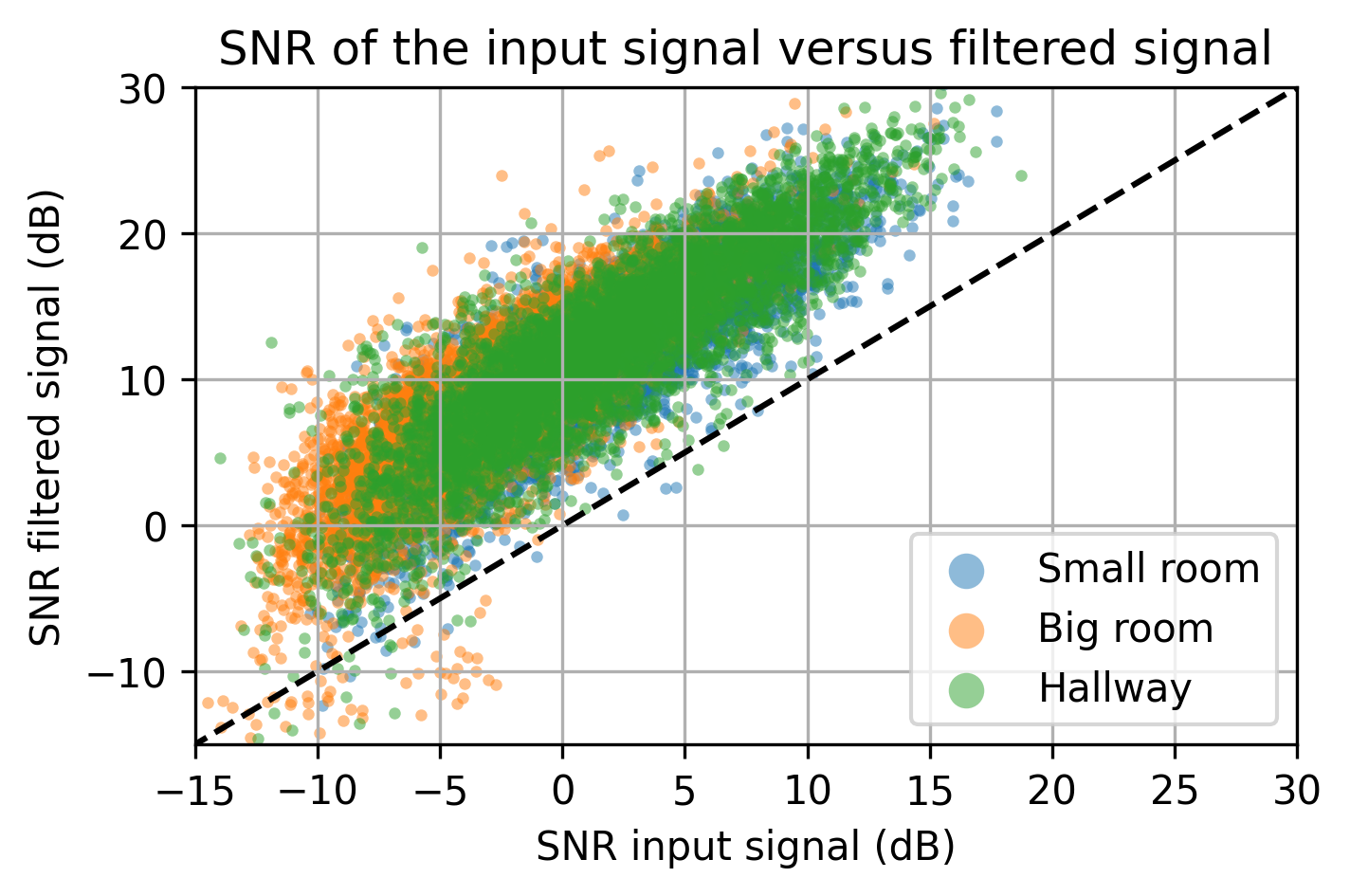}
\vspace{-20pt}
\caption{SNRs before and after filtering on all 0.5 sec segments}
\label{fig_snr}
\vspace{-5pt}
\end{figure}

\begin{figure}
\centering
\includegraphics[width=\linewidth]{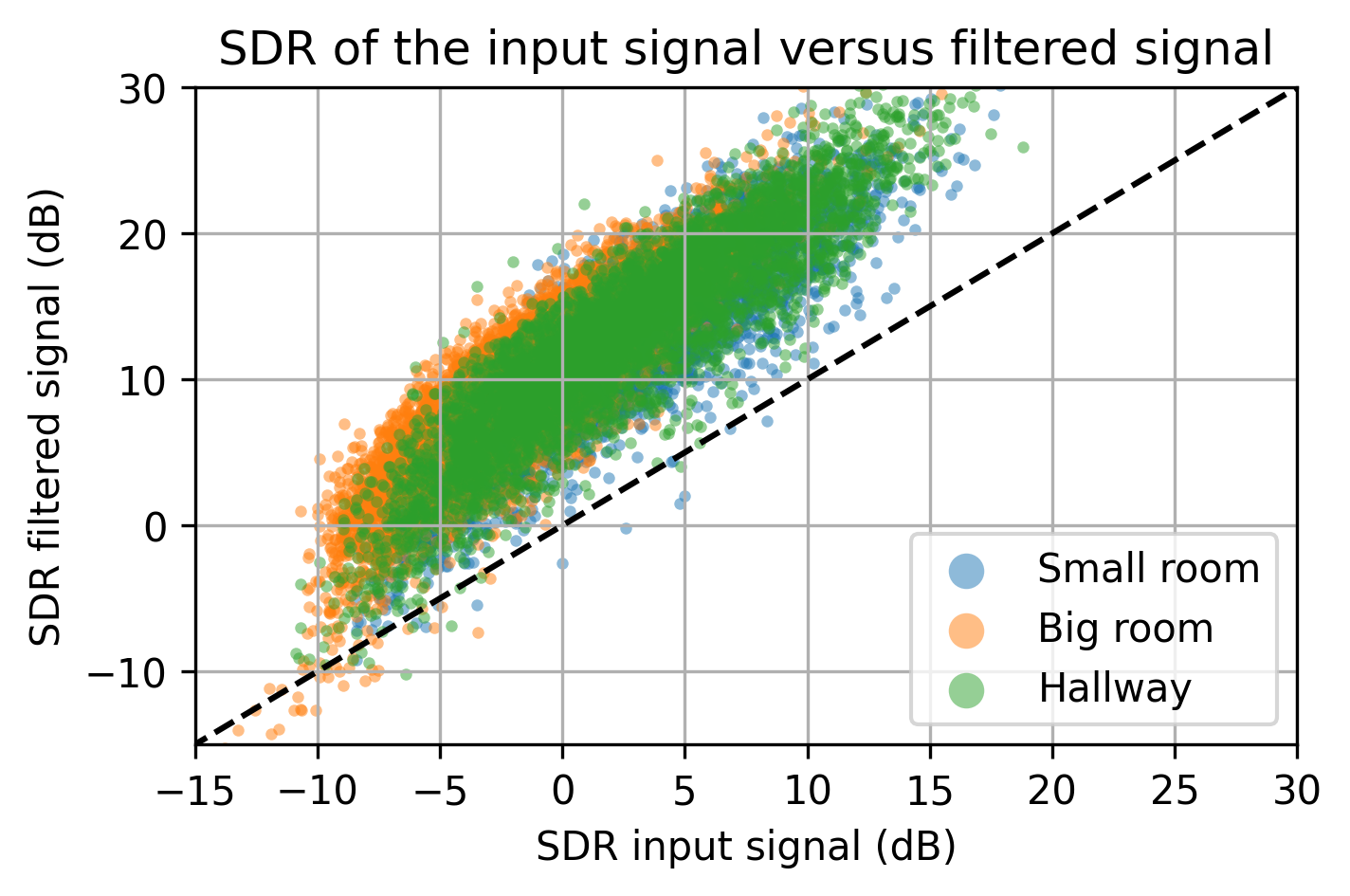}
\vspace{-20pt}
\caption{SDRs before and after filtering on all 0.5 sec segments}
\label{fig_sdr}
\end{figure}

Figures \ref{fig_snr} and \ref{fig_sdr} show scatter plots of the input and filtered SNRs and SDRs for all 0.5 sec segments.
The results shows that the enhanced signal have a better SNR and SDR compared to the input signal.
Results show linear performances, except at low SNRs (around -5 and -10 dB), where the weak power of the target sounds in the audio signal can cause deterioration of the MVDR performances.

Table \ref{table_wer} shows the average WER of the input, enhanced and voice only signal. 
As expected, the WER of the input signal in all three rooms is higher compared of the voice-only signal. 
The proposed ego-noise reduction method reduces considerably the WER by 30\% to 55\%. 
Input signals recorded in the small room and the hallway show a higher SNR, resulting in a better WER results with the input signal. 

\begin{table}
\caption{Word error rate (WER) with the proposed framework}
\vspace{-10pt}
\label{table_wer}
\begin{center}
\renewcommand{\arraystretch}{1.5}
\begin{tabular}{c c c c}
\hline
\hline
\textbf{Room} & \textbf{Input} & \textbf{Enhanced} & \textbf{Voice only}\\
\hline
Small & 62.6\% & 30.9\% & 11.1\%\\
Large & 90.3\% & 35.5\% & 14.9\%\\
Hallway & 62.4\% & 25.3\% & 8.52\%\\
\hline
\hline
\end{tabular}
\end{center}
\end{table}

\subsection{Sound event detection}
\label{sec:results_event}

\begin{figure}
    \centering
    \includegraphics[width=\linewidth]{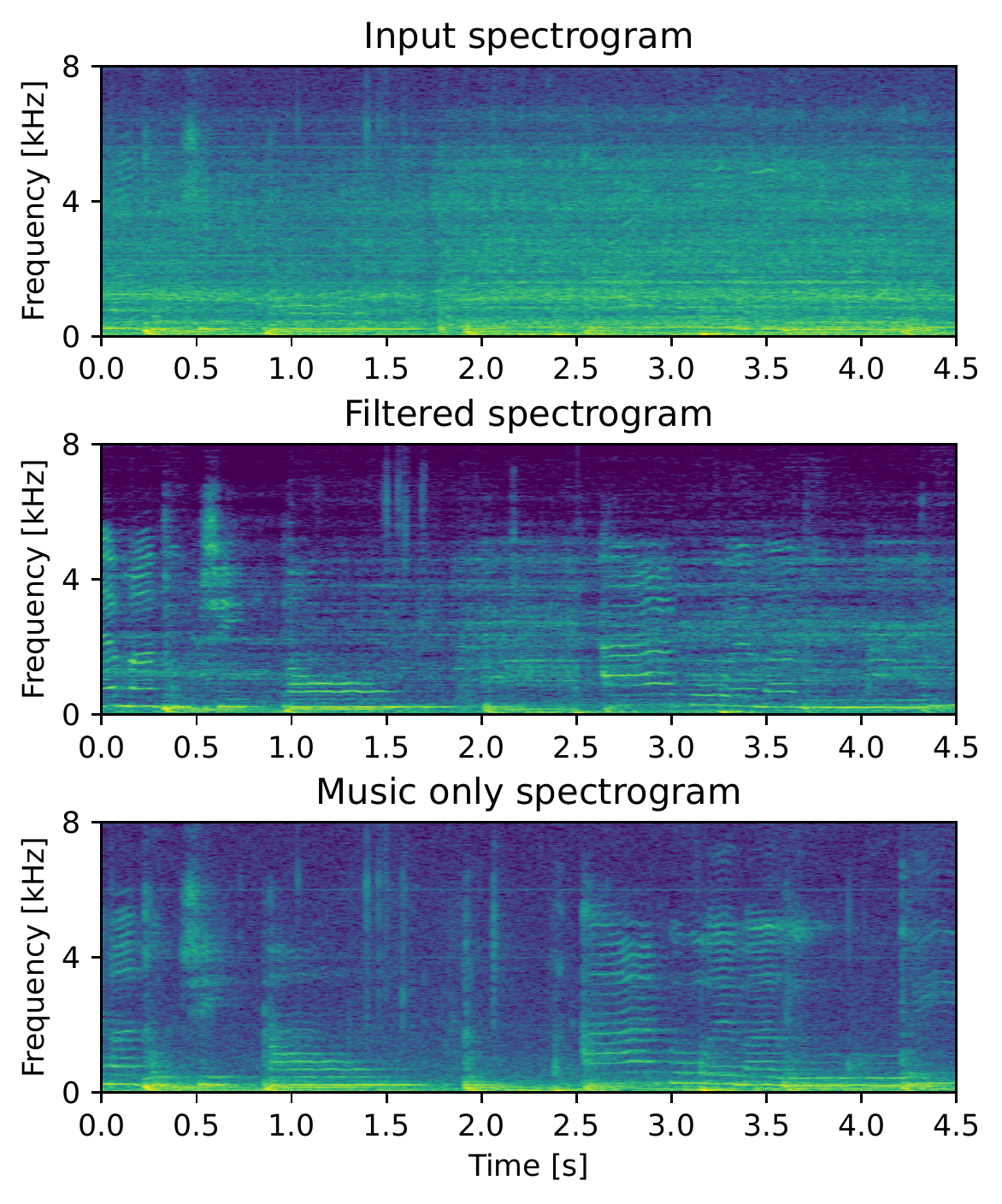}
    \vspace{-20pt}
    \caption{Spectrograms on the input, filtered and music only signal. The SNR of the input and filered signal are -1.93 dB and 13.13 dB, respectively.}
    \label{fig:specs_event}
\end{figure}

Figure \ref{fig:specs_event} shows an example of the input, filtered and music-only spectrograms in a large room with SNRs of $\text{-1.93}$ dB for the input signal and 13.13 dB for the filtered signal.
While music is mainly masked by ego-noise in the input signal spectrogram, harmonics become visible in the enhanced signal spectrogram. 

\begin{figure}
\centering
\includegraphics[width=\linewidth]{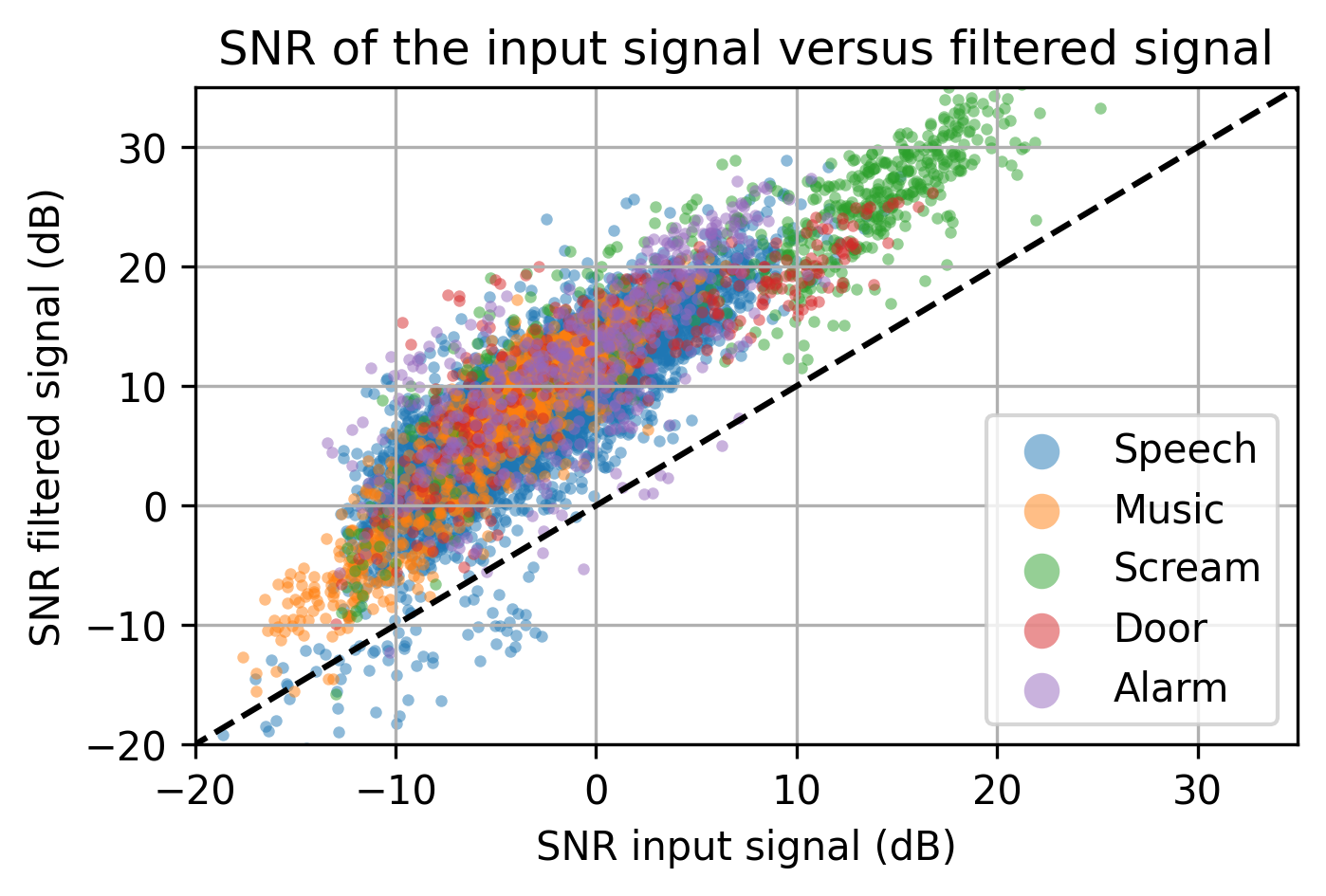}
\vspace{-20pt}
\caption{SNRs before and after filtering for each sound event type}
\label{fig_snr_event}
\end{figure}

\begin{figure}
\centering
\includegraphics[width=\linewidth]{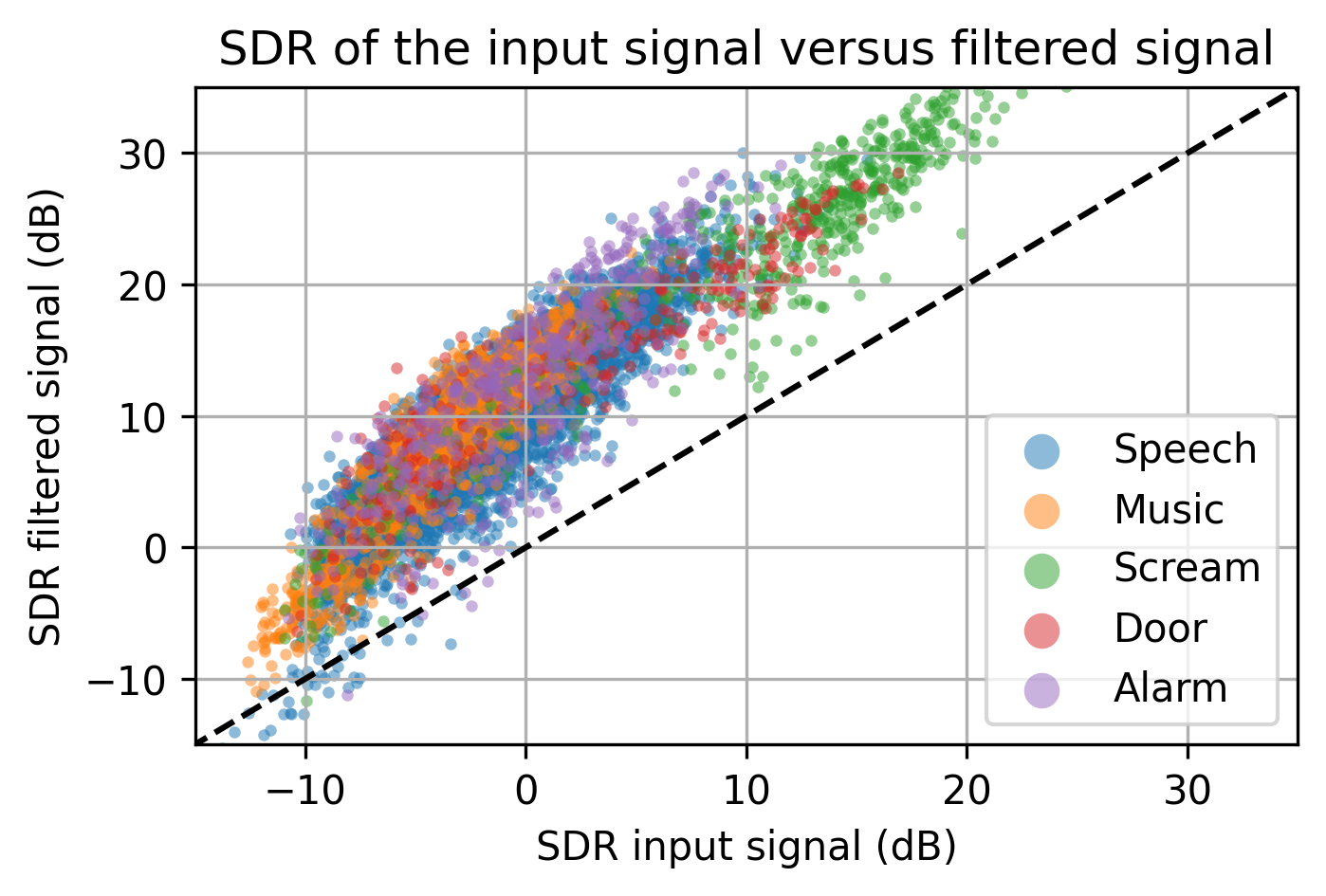}
\vspace{-20pt}
\caption{SDRs before and after filtering for each sound event type}
\label{fig_sdr_event}
\end{figure}

Figures \ref{fig_snr_event} and \ref{fig_sdr_event}  show the input and filtered SNR and SDR for all the events in the large room. Since results are similar in all three rooms, we focus on the large room for clarity.  
While the difference between each event is mainly the SNR of the input signal, the system improves the sound quality in most scenarios.

The PANNS model is used to classify the sound events using the input, enhanced and event only signals. 
Results shown in Table \ref{table_event} suggest that the enhanced signal increase the AP on music signal.
Some results show a smaller AP in event only signal than noisy signal and even bigger with the enhanced signal.
Results are mainly better than observed in \cite{kong2020panns}.
Unlike in Audioset, data recorded in this experiment is clear and isolate from other sound.
Mitigated performances are however observed with the door slam sound event.
We believe this to be be due to the short and faint sounds of a door closing, which can be more difficult to capture when computing the input signal spatial covariance matrix.
While the alarm sound event remains difficult to detect with PANNS, the framework still provide some slight increase for the AP.
It is believed that the framework is more beneficial to audio segment with lower SNR values.

\begin{table}
\caption{Average precision of event detection with the proposed framework}
\vspace{-10pt}
\label{table_event}
\begin{center}
\renewcommand{\arraystretch}{1.5}
\begin{tabular}{c c c c c}

\hline
\hline

\textbf{Event} & \textbf{Room} & \textbf{Input} & \textbf{Enhanced} & \textbf{Event only}\\
\hline
\multirow{3}{4em}{Speech} & Small & 0.999 & 1.0 & 1.0\\
& Large & 0.994 & 1.0 & 1.0\\
& Hallway & 0.999 & 1.0 & 1.0\\
\hline
\multirow{3}{4em}{Music} & Small & 0.722 & 0.792 & 0.914\\
& Large & 0.487 & 0.685 & 0.784\\
& Hallway & 0.868 & 0.953 & 0.948\\
\hline
\multirow{3}{4em}{Scream} & Small & 0.993 & 0.994 & 0.992\\
& Large & 0.998 & 0.998 & 1.0\\
& Hallway & 0.994 & 1.0 & 1.0\\
\hline
\multirow{3}{4em}{Door} & Small & 0.313 & 0.232 & 0.104\\
& Large & 0.178 & 0.109 & 0.091\\
& Hallway & 0.129 & 0.100 & 0.089\\
\hline
\multirow{3}{4em}{Alarm} & Small & 0.091 & 0.105 & 0.108\\
& Large & 0.091 & 0.093 & 0.096\\
& Hallway & 0.090 & 0.098 & 0.105\\
\hline
\hline
\end{tabular}
\end{center}
\vspace{-10pt}
\end{table}

\section{CONCLUSIONS}
This paper introduce an efficient framework to reduce the ego-noise of a robot using a microphones array. 
The framework relies on a dictionary of noise SCMs trained during a short calibration period and the MVDR algorithms to enhanced the target sounds sources. 
For better performance, the SCMs are concatenated in a supervector, which dimensions are reduced offline using the PCA algorithm.
Results obtained with a UGV and a 16-microphone array show an improvement of approximately 10 dB on average for the SNRs and SDRs in the enhanced signal using only 90 sec of calibration time. 
Moreover, once the enhanced speech signal is fed to a automatic recognition system, the word error rate decreases by to 55\% compared to the performances obtained using the noisy input signals.
This approach also improves average precision of sound event detection performances, by up to 0.2 in some cases.
This new framework bring new opportunities to integrate speech recognition and sound sources detection on robots that generate unseen ego-noise at training time for deep neural network models.






\section*{ACKNOWLEDGMENT}
The authors would like to thank Prof. Alexandre Girard for providing access to a Clearpath Robotics Jackal robot, and Samuel Faucher for the insightful discussions.


\bibliographystyle{IEEEtran}
\bibliography{IEEEabrv, references}

\begin{thebibliography}{10}
\providecommand{\url}[1]{#1}
\csname url@rmstyle\endcsname
\providecommand{\newblock}{\relax}
\providecommand{\bibinfo}[2]{#2}
\providecommand\BIBentrySTDinterwordspacing{\spaceskip=0pt\relax}
\providecommand\BIBentryALTinterwordstretchfactor{4}
\providecommand\BIBentryALTinterwordspacing{\spaceskip=\fontdimen2\font plus
\BIBentryALTinterwordstretchfactor\fontdimen3\font minus
  \fontdimen4\font\relax}
\providecommand\BIBforeignlanguage[2]{{%
\expandafter\ifx\csname l@#1\endcsname\relax
\typeout{** WARNING: IEEEtran.bst: No hyphenation pattern has been}%
\typeout{** loaded for the language `#1'. Using the pattern for}%
\typeout{** the default language instead.}%
\else
\language=\csname l@#1\endcsname
\fi
#2}}

\bibitem{radford2022robust}
A.~Radford, J.~W. Kim, T.~Xu, G.~Brockman, C.~McLeavey, and I.~Sutskever,
  ``Robust speech recognition via large-scale weak supervision,'' \emph{arXiv
  preprint arXiv:2212.04356}, 2022.

\bibitem{chiu2018state}
C.-C. Chiu, T.~N. Sainath, Y.~Wu, R.~Prabhavalkar, P.~Nguyen, Z.~Chen,
  A.~Kannan, R.~J. Weiss, K.~Rao, E.~Gonina, \emph{et~al.}, ``State-of-the-art
  speech recognition with sequence-to-sequence models,'' in \emph{Proceedings
  of the IEEE International Conference on Acoustics, Speech and Signal
  Processing}, 2018, pp. 4774--4778.

\bibitem{kong2020panns}
Q.~Kong, Y.~Cao, T.~Iqbal, Y.~Wang, W.~Wang, and M.~D. Plumbley, ``{PANNS}:
  Large-scale pretrained audio neural networks for audio pattern recognition,''
  \emph{IEEE/ACM Transactions on Audio, Speech, and Language Processing},
  vol.~28, pp. 2880--2894, 2020.

\bibitem{wang_tfs_2017}
L.~Wang and A.~Cavallaro, ``Time-frequency processing for sound source
  localization from a micro aerial vehicle,'' in \emph{Proceedings of the
  {IEEE} International Conference on acoustics, Speech and Signal Processing},
  2017, pp. 496--500.

\bibitem{wang_acoustic_2018}
------, ``Acoustic sensing from a multi-rotor drone,'' \emph{{IEEE} Sensors
  Journal}, vol.~18, no.~11, pp. 4570--4582, 2018.

\bibitem{wang_ear_2016}
------, ``Ear in the sky: Ego-noise reduction for auditory micro aerial
  vehicles,'' in \emph{Proceedings of the {IEEE} International Conference on
  Advanced Video and Signal Based Surveillance}, 2016, pp. 152--158.

\bibitem{wang_microphone-array_2017}
------, ``Microphone-array ego-noise reduction algorithms for auditory micro
  aerial vehicles,'' \emph{{IEEE} Sensors Journal}, vol.~17, no.~8, pp.
  2447--2455, 2017.

\bibitem{wang_blind_2020}
------, ``A blind source separation framework for ego-noise reduction on
  multi-rotor drones,'' \emph{{IEEE}/{ACM} Transactions on Audio, Speech, and
  Language Processing}, vol.~28, pp. 2523--2537, 2020.

\bibitem{kang_software_2019}
B.~Kang, H.~Ahn, and H.~Choo, ``A software platform for noise reduction in
  sound sensor equipped drones,'' \emph{{IEEE} Sensors Journal}, vol.~19,
  no.~21, pp. 10\,121--10\,130, 2019.

\bibitem{yoon_advanced_2015}
S.~Yoon, S.~Park, Y.~Eom, and S.~Yoo, ``Advanced sound capturing method with
  adaptive noise reduction system for broadcasting multicopters,'' in
  \emph{Proceedings of the {IEEE} International Conference on Consumer
  Electronics}, 2015, pp. 26--29.

\bibitem{yoon_two-stage_2016}
S.~Yoon, S.~Park, and S.~Yoo, ``Two-stage adaptive noise reduction system for
  broadcasting multicopters,'' in \emph{Proceedings of the {IEEE} International
  Conference on Consumer Electronics}, 2016, pp. 219--222.

\bibitem{fernandes_first_2015}
R.~P. Fernandes, E.~C. Santos, A.~L.~L. Ramos, and J.~A. Apolinário~Jr., ``A
  first approach to signal enhancement for quadcopters using piezoelectric
  sensors,'' in \emph{Proceedings of the International Conference on
  Transformative Science and Engineering, Business and Social Innovation},
  2015, pp. 536--541.

\bibitem{ince_online_2012}
G.~Ince, K.~Nakadai, and K.~Nakamura, ``Online learning for template-based
  multi-channel ego noise estimation,'' in \emph{Proceedings of the
  {IEEE}/{RSJ} International Conference on Intelligent Robots and Systems},
  2012, pp. 3282--3287.

\bibitem{ince_ego_2009}
G.~Ince, K.~Nakadai, T.~Rodemann, Y.~Hasegawa, H.~Tsujino, and J.~Imura, ``Ego
  noise suppression of a robot using template subtraction,'' in
  \emph{Proceedings of the {IEEE}/{RSJ} International Conference on Intelligent
  Robots and Systems}, 2009, pp. 199--204.

\bibitem{ince_hybrid_2010}
G.~{Ince}, K.~Nakadai, T.~Rodemann, Y.~Hasegawa, H.~Tsujino, and J.~Imura, ``A
  hybrid framework for ego noise cancellation of a robot,'' in
  \emph{Proceedings of the {IEEE} International Conference on Robotics and
  Automation}, 2010, pp. 3623--3628.

\bibitem{ince_incremental_2011}
G.~Ince, K.~Nakadai, T.~Rodemann, J.~Imura, K.~Nakamura, and H.~Nakajima,
  ``Incremental learning for ego noise estimation of a robot,'' in
  \emph{Proceedings of the {IEEE}/{RSJ} International Conference on Intelligent
  Robots and Systems}, 2011, pp. 131--136.

\bibitem{schmidt_informed_2019}
A.~Schmidt and W.~Kellermann, ``Informed ego-noise suppression using motor
  data-driven dictionaries,'' in \emph{Proceedings of the {IEEE} International
  Conference on Acoustics, Speech and Signal Processing}, 2019, pp. 116--120.

\bibitem{schmidt_motor_2020}
A.~Schmidt, A.~Brendel, T.~Haubner, and W.~Kellermann, ``Motor data-regularized
  nonnegative matrix factorization for ego-noise suppression,'' \emph{Journal
  on Audio, Speech, and Music Processing}, vol.~11, pp. 1--15, 2020.

\bibitem{schmidt_multichannel_2021}
A.~Schmidt and W.~Kellermann, ``Multichannel nonnegative matrix factorization
  with motor data-regularized activations for robust ego-noise suppression,''
  in \emph{Proceedings of the {IEEE} International Conference on Autonomous
  Systems}, 2021, pp. 1--5.

\bibitem{deleforge_phase-optimized_2015}
A.~Deleforge and W.~Kellermann, ``Phase-optimized k-{SVD} for signal extraction
  from underdetermined multichannel sparse mixtures,'' in \emph{Proceedings of
  the {IEEE} International Conference on Acoustics, Speech and Signal
  Processing}, 2015, pp. 355--359.

\bibitem{tezuka_ego-motion_2014}
T.~Tezuka, T.~Yoshida, and K.~Nakadai, ``Ego-motion noise suppression for
  robots based on semi-blind infinite non-negative matrix factorization,'' in
  \emph{Proceedings of the {IEEE} International Conference on Robotics and
  Automation}, 2014, pp. 6293--6298.

\bibitem{haubner_multichannel_2018}
T.~Haubner, A.~Schmidt, and W.~Kellermann, ``Multichannel nonnegative matrix
  factorization for ego-noise suppression,'' in \emph{Proceedings of the Speech
  Communication}, 2018, pp. 136--140.

\bibitem{fang_joint_2021}
H.~Fang, G.~Carbajal, S.~Wermter, and T.~Gerkmann, ``Joint reduction of
  ego-noise and environmental noise with a partially-adaptive dictionary,'' in
  \emph{Proceedings of the {ITG} Conference on Speech Communication}, 2021, pp.
  114--118.

\bibitem{ito_internal_2005}
A.~Ito, T.~Kanayama, M.~Suzuki, and S.~Makino, ``Internal noise suppression for
  speech recognition by small robots.'' in \emph{Proceedings of the European
  Conference on Speech Communication and Technology}, 2005, pp. 2685--2688.

\bibitem{briegleb_deep_2019}
A.~Briegleb, A.~Schmidt, and W.~Kellermann, ``Deep clustering for
  single-channel ego-noise suppression,'' in \emph{Proceedings of the
  International Congress on Acoustics}, 2019, pp. 2813--2820.

\bibitem{kim_mimo_2021}
J.~Kim, J.~Choi, J.~Son, G.~Kim, J.~Park, and J.~Chang, ``{MIMO} noise
  suppression preserving spatial cues for sound source localization in mobile
  robot,'' in \emph{Proceedings of the {IEEE} International Symposium on
  Circuits and Systems}, 2021, pp. 1--5.

\bibitem{tan_efficient_2019}
Z.~Tan, A.~H.~T. Nguyen, and A.~W.~H. Khong, ``An efficient dilated
  convolutional neural network for {UAV} noise reduction at low input {SNR},''
  in \emph{Proceedings of the Asia-Pacific Signal and Information Processing
  Association Annual Summit and Conference}, 2019, pp. 1885--1892.

\bibitem{wang_deep_2020}
L.~Wang and A.~Cavallaro, ``Deep learning assisted time-frequency processing
  for speech enhancement on drones,'' \emph{{IEEE} Transactions on Emerging
  Topics in Computational Intelligence}, vol.~5, no.~6, pp. 871--881, 2020.

\bibitem{panayotov2015librispeech}
V.~Panayotov, G.~Chen, D.~Povey, and S.~Khudanpur, ``Librispeech: an {ASR}
  corpus based on public domain audio books,'' in \emph{Proceedings of the IEEE
  International Conference on Acoustics, Speech and Signal Processing}, 2015,
  pp. 5206--5210.

\end{thebibliography}

\end{document}